\documentclass[twocolumn]{aastex701}
\usepackage{tabularx}
\usepackage{amsmath}
\usepackage{graphicx}
\usepackage{booktabs}
\usepackage{multirow}
\usepackage{accents}
\usepackage{tablefootnote}
\let\tablenum\relax 
\usepackage{siunitx}
\usepackage{needspace}  \Needspace{5\baselineskip}
\usepackage{placeins}
\usepackage{longtable}
\newcommand{\tess}{\textit{TESS}}
\newcommand{\name}{KELT-20}
\newcommand{\pname}{\texorpdfstring{KELT-20\,b}{KELT-20\,b}}
\newcommand\RJ{$R_{\mathrm{J}}$}
\newcommand\MJ{$M_{\mathrm{J}}$}

\newcommand{\totaltd}{19} 
\newcommand{\totalpm}{77} 

\newcommand{\age}{$58\pm5$\,Myr}

\newcommand{\mass}{$<3.4$}
\newcommand{\dist}{138} 
\newcommand{\porb}{3.5}

\shortauthors{Distler et al.}
\begin{document}

\title{TESS Hunt for Young and Maturing Exoplanets (THYME) XIV: \\ 
A Comoving-Based Age Constraint for KELT-20}

\author[orcid=0009-0006-4294-6760]{Adam Distler}
\altaffiliation{NSF Graduate Research Fellow}
\affiliation{Center for Astrophysics | Harvard and Smithsonian, 60 Garden Street, Cambridge, MA 02138, USA}
\affiliation{Department of Astronomy, University of Wisconsin-Madison, 475 N.~Charter St., Madison, WI 53706, USA}
\affiliation{Wisconsin Center for Origins Research, University of Wisconsin--Madison, 6515 Sterling Hall, 475 N.
Charter St., Madison, WI, USA, 53706}
\email[show]{adamdistler@cfa.harvard.edu}

\author[orcid=0000-0001-7493-7419]{Melinda Soares-Furtado}
\affiliation{Department of Astronomy, University of Wisconsin-Madison, 475 N.~Charter St., Madison, WI 53706, USA}
\affiliation{Department of Physics \& Kavli Institute for Astrophysics and Space Research, Massachusetts Institute of Technology, Cambridge, MA 02139, USA}
\affiliation{Wisconsin Center for Origins Research, University of Wisconsin--Madison, 6515 Sterling Hall, 475 N.
Charter St., Madison, WI, USA, 53706}
\email{mmsoares@wisc.edu}

\author[orcid=0000-0003-3654-1602]{Andrew W.~Mann}
\affiliation{Department of Physics \& Astronomy, The University of North Carolina at Chapel Hill, Chapel Hill, NC 27599-3255, USA}
\email{awmann@unc.edu}

\author[orcid=0000-0001-9811-568X]{Adam L.~Kraus}
\affiliation{Department of Astronomy, The University of Texas at Austin, Austin, TX 78712, USA}
\email{alk@astro.as.utexas.edu}

\author[orcid=0000-0002-2592-9612]{Jonathan Gagn\'e}
\affiliation{Plan\'etarium de Montr\'eal, Espace pour la Vie, 4801 av. Pierre-de Coubertin, Montr\'eal, Qu\'ebec, Canada}
\affiliation{Trottier Institute for Research on Exoplanets, Universit\'e de Montr\'eal, D\'epartement de Physique, C.P.~6128 Succ. Centre-ville, Montr\'eal, QC H3C~3J7, Canada}
\email{jonathan.gagne@montreal.ca}

\author[orcid=0000-0002-7733-4522]{Juliette Becker}
\affiliation{Department of Astronomy, University of Wisconsin-Madison, 475 N.~Charter St., Madison, WI 53706, USA}
\email{juliette.becker@wisc.edu}

\author[orcid=0009-0007-0488-5685]{Ritvik Sai Narayan}
\affiliation{Department of Astronomy, University of Wisconsin-Madison, 475 N.~Charter St., Madison, WI 53706, USA}
\email{rnarayan4@wisc.edu}

\author[orcid=0009-0009-8236-0862]{Max Clark}
\affiliation{Department of Astronomy, University of Wisconsin-Madison, 475 N.~Charter St., Madison, WI 53706, USA}
\affiliation{Department of Astronomy, University of Michigan, Ann Arbor, MI 48109, USA}
\email{mclark.astro@gmail.com}

\author[orcid=0000-0001-7246-5438]{Andrew Vanderburg}
\affiliation{Center for Astrophysics | Harvard and Smithsonian, 60 Garden Street, Cambridge, MA 02138, USA}
\email{avanderburg@cfa.harvard.edu}

\author[orcid=0000-0001-8812-0565]{Joseph E.~Rodriguez} 
\affiliation{Center for Data Intensive and Time Domain Astronomy, Department of Physics and Astronomy, Michigan State University, East Lansing, MI 48824, USA}
\email{jrod@msu.edu}

\author[orcid=0000-0002-3553-9474]{Laura K.~Rogers}
\affiliation{NOIRLab, 950 N Cherry Ave, Tucson, AZ, 85719, USA}
\email{laura.rogers@noirlab.edu}

\author[orcid=0000-0002-6549-9792]{Ronan Kerr}
\affiliation{Dunlap Institute for Astronomy \& Astrophysics, University of Toronto
Toronto, ON M5S 3H4, Canada}
\email{ronan.kerr@utoronto.ca}

\correspondingauthor{Adam Distler}

\begin{abstract}
Young stellar moving groups offer unique opportunities to investigate the early evolution of stellar and planetary systems. In continuation of an ongoing effort to age-date compelling planetary systems, we provide an in-depth age analysis of \name{}---a young A-type star that hosts a well-aligned ultra-hot Jupiter. This system poses a useful case study to investigate migration mechanisms at early stages of evolution.
Using \textit{Gaia} DR3 data, we identify \totalpm{} stars with proper motions consistent with \name{}, including \totaltd{} with measured radial velocities that enable full 3D kinematic confirmation. Using isochronal analyses, gyrochronology, photometric variability, and stellar activity indicators, we converge on an age of \age{}. This constraint provides critical insights into the dynamical processes shaping hot Jupiter formation.
\end{abstract}


\section{Introduction}\label{sec:intro}
Young stellar systems serve as valuable laboratories to investigate planetary evolution during formative stages. These systems probe atmospheric loss \citep[e.g.,][]{Lammer2003, Lopez2013,Ginzburg2018}, bulk property evolution \citep[e.g.][]{Fortney2011, Krumholz2019}, and dynamical evolution \citep[e.g.,][]{1998Icar..136..304C,Asphaug+2006, Fabrycky2007, Chatterjee+2008, Ford+2008, Hansen2009, Naoz2011,2012AREPS..40..251M}.
Determining accurate and precise planetary ages is a cornerstone of these efforts. The ages of exoplanets are typically inferred from their host stars, under the assumption that stars and their planets form coevally. 
Age-dating techniques are especially powerful when applied to stars in young moving groups, associations, and stellar clusters.
This stems from the fact that stars in these groups originate from the same molecular cloud, sharing a common age and chemical composition.
The ages of the stars within such systems can be evaluated as an ensemble, using reliable tools like isochronal fits and color-rotation (gyrochronological) sequences \citep{Soderblom2010}. 
Furthermore, these environments often provide multiple independent lines of evidence that can be evaluated to ensure the reliability of the age estimates. 

In this work, we present a stellar association-based age constraint for \name{} (MASCARA-2, \textit{Gaia} DR3 2033123654092592384, TOI-1151, TIC 69679391). This is a young \citep{KounkelCovey2020, Talens2018}, bright ($V = 7.6$\,mag) A-type star, hosting a well-aligned ultra-hot Jupiter (\citealt{Lund2017,Talens2018}). We aim to provide an independent and more comprehensive study of the comoving targets to \name{} to refine prior age estimates. This work builds on current efforts to identify and characterize stellar associations \citep[e.g.,][]{Oh2017, Kounkel2019, Newton2019, Moranta2022}, enabled by the abundance and quality of astrometric data provided by the \textit{Gaia} satellite \citep{GaiaMission}.

We made use of isochronal fits, gyrochronology, and photometric variability to obtain a multifaceted age estimate, converging on an age of \age{}.
\name{} provides a valuable test case for exploring the formation and dynamical evolution of planetary systems in young environments with well-constrained ages.

In Section~\ref{sec:materialsmethods}, we describe the data products and methods used to determine which sources are comoving with \name{}, as well as four independent analyses of the system age.
Section~\ref{sec:comparison} compares this work with past age analyses of \name{}.
Section~\ref{sec:Kelt-20b} describes the age implications for the planetary companion and the potential for informative follow-up investigations. 
In Section~\ref{sec:summary}, we summarize our findings.

\section{Data \& Methods}\label{sec:materialsmethods}

\subsection{Identification of a Comoving Sample}\label{sec:Comove}

To identify comoving targets to \name{}, we made use of the publicly-available \texttt{Python} routine \texttt{FriendFinder} \citep{toffelmire2021}. 
Upon specifying a given target, the algorithm employs parallaxes, PMs, and radial velocities from the \textit{Gaia} DR3 survey \citep{GaiaMission, GaiaCollaboration2021} to find comoving companions to a given target. In short, this algorithm calculates the UVW galactic velocity values given a target of interest, and then projects them into sky-plane and radial velocities for every other target within a given search radius from that target of interest. These values are then compared to those measured by \textit{Gaia} DR3. For a full description of the algorithm, consult \cite{toffelmire2021}.

We used \name{} as our initial target, specifying an RV of $-23.3$\,km\,s$^{-1}$ \citep{Lund2017}. We opted not to use the \textit{Gaia} DR3 \citep{gaiadr3} RV of $-26.78\,\mathrm{km}\ \mathrm{s}^{-1}$ given the superior spectroscopic resolution ($\Delta \lambda/\lambda\approx 44,000$) and wavelength range  (3900-9100\,\AA) of the TRES spectrograph \citep{Szentgyorgyi2007, Furesz08} compared to that of the \textit{Gaia} Radial Velocity Spectrometer ($\Delta\lambda /\lambda = 11,500$ and 8450-8720\,\AA; \citealt{Katz2023}). Broader spectral coverage is advantageous for characterizing the rapidly rotating nature of \name{} ($v\sin(i)=114$\,km\,s$^{-1}$, \citealt{Lund2017}), as the spectrum is dominated by hydrogen lines.

We incorporated a 40\,pc search radius and removed all stars with sky-plane tangential velocities that differed by more than 1\,km\,s$^{-1}$ from the predicted value for a comoving source. This resulted in a list of 131 stars.

Further, to ensure the fidelity of our sample, we elected to remove binaries from our comoving target list, as binarity can: (a) induce shifts in radial velocities on the order of tens of kilometers per second, leading to uncertainty of true 3D comovement; (b) shift color-magnitude diagram position, providing erroneous isochronal fits; and (c) impact the angular momentum history of stars, compromising gyrochronology-based age estimates. 
To exclude likely binary systems, we applied an upper limit of 1.25 on the \textit{Gaia} DR3 renormalized unit weight error (RUWE; \citealt{belokurov, Penoyre2022}) to both the PM and 3D kinematic samples. This resulted in a sample of 106 comoving proper motion candidates.

\begin{figure*}[htb!]
    \centering
    \includegraphics[width=0.85\linewidth]{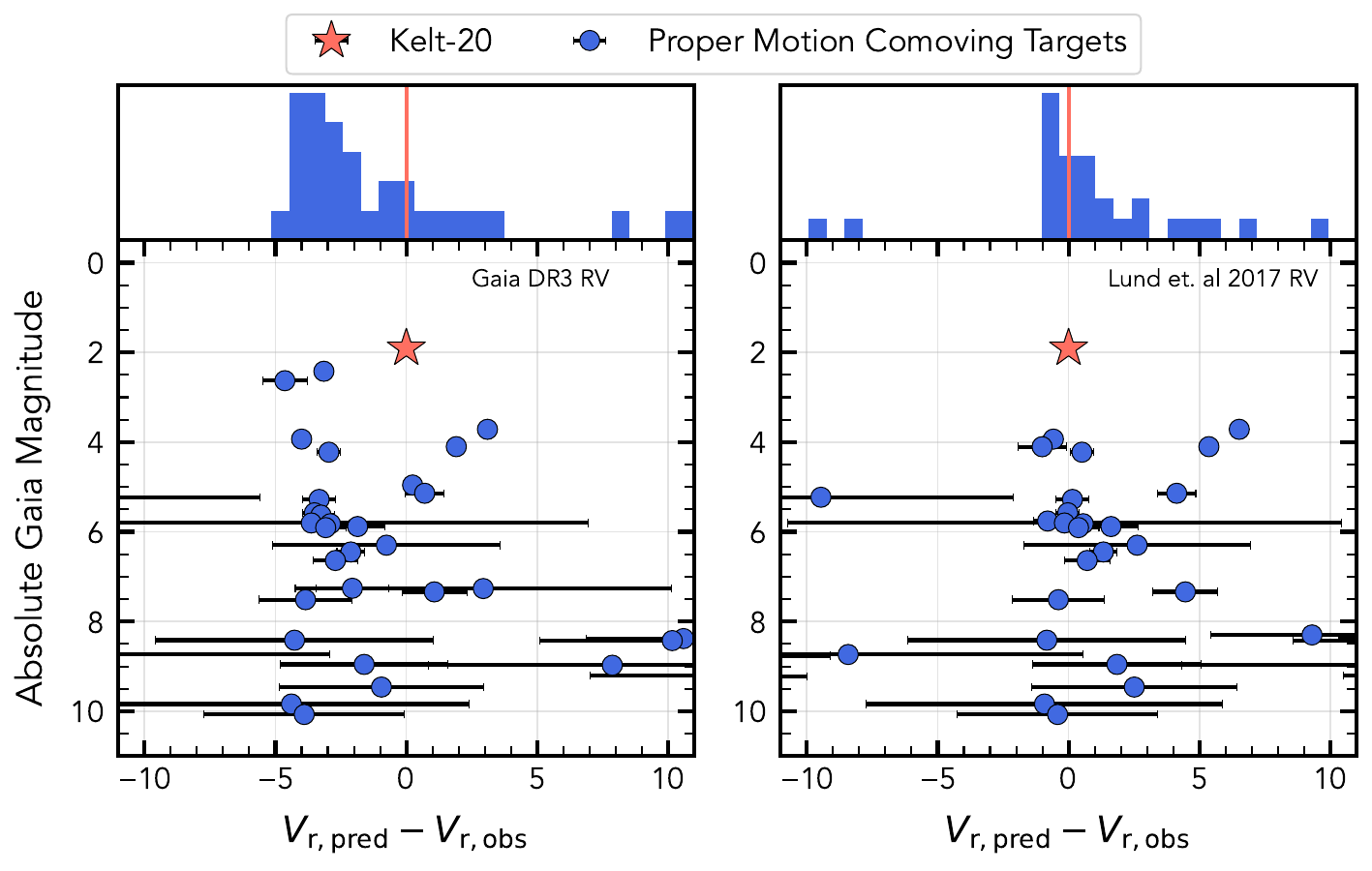}
    \caption{Comparison of \texttt{FriendFinder} results obtained using the 
    \textit{Gaia} DR3 radial velocity (left) and the \citet{Lund2017} radial velocity (right) for \name{}. 
    Plotted are the differences between predicted ($V_\mathrm{r,pred}$) and observed ($V_\mathrm{r,obs}$) radial velocities as a function of absolute \textit{Gaia} magnitude. 
    Blue circles mark stars with RUWE~$<1.25$. 
    The \citet{Lund2017} RV yields a closer alignment with the overdensity, supporting its adoption in this analysis.}
    \label{fig:LundvGaia}
\end{figure*}

To establish true 3D comovement, we then compared the observed \textit{Gaia} DR3 RV measurements for each candidate to the algorithm-predicted values using the \cite{Lund2017} RV. For the PM sources not possessing a \textit{Gaia} DR3 RV, we searched for RVs in the Pulkovo Compilation of Radial Velocities \citep{Gontcharav2006}.
The \texttt{FriendFinder} comparison between the \cite{Lund2017} RV and the \textit{Gaia} DR3 RV input into \texttt{FriendFinder} is shown in Figure~\ref{fig:LundvGaia}.
The \cite{Lund2017} RV clearly provides a closer alignment with the overdensity apparent in the plot.
We then removed targets with RV offsets greater than 3\,km\,s$^{-1}$ from the value predicted by the \texttt{FriendFinder} algorithm. 
This resulted in a population of \totaltd{} 3D kinematic comoving targets.

\subsection{Investigation of Chance Kinematic Agreement}
To evaluate the extent to which our comoving targets reflect the properties of a coherent moving group, we compared the galactic velocity $\langle U,V,W\rangle$ values against the distribution of all stars in the 40\,pc search radius. 
A genuine overdensity in $\langle U,V,W\rangle$ space would strengthen confidence in the validity of the system. Within this search volume, we identified 5516 stars in \textit{Gaia} DR3 (including our \totaltd{} 3D comoving targets) that possess full spatial and kinematic information and satisfy the binarity cut of RUWE~$<1.25$. The velocity distribution extends across several hundred km\,s$^{-1}$.

To assess the likelihood of field star contamination in the identified comoving group, we performed a Monte Carlo simulation assuming that the distribution of stars within the \texttt{FriendFinder} search radius follows a multivariate Gaussian in $\langle U, V, W \rangle$ velocity space. We drew random samples from this distribution, each containing 5516 stars (the number of sources within 40\,pc of \name{} with complete kinematic and spatial information) and repeated this process 5000 times. For each realization, we counted how many stars fell within the $\langle U, V, W \rangle$ bounds of the comoving targets, representing the number of interlopers that could be expected by chance. Across the ensemble of trials, we found a mean contamination level of $0^{+1}_{-0}$ stars. This implies that, for the full sample of \totaltd{} 3D comoving targets, $0 \pm 5$\% are statistically expected to be field star interlopers. These results indicate that the kinematic volume surrounding \name{} is significantly overdense in velocity space relative to random field stars, strengthening the case that it represents a coeval population.

\subsection{Isochronal Age Estimation}
\label{sec:Siesta}
Determining isochronal ages of main-sequence (MS) stars is particularly challenging, since they spend the majority of their lifetimes in this phase and their positions on a color--magnitude diagram (CMD) change very little. By identifying comoving targets to \name{}, we can leverage key age indicators on the CMD such as the MS turn-off and pre-Main-Sequence (pMS) population to constrain \name{}'s age.

To estimate an isochronal age and associated error for our group of comoving stars, we employed the mixture model from \cite{Mann2022}, which is based on the statistical models from \cite{Hogg2010}. In short, the model synthesizes two populations: a coeval population, drawn from a \texttt{PARSEC 1.2S} isochrone \citep{Bressan2012} from reddening ($E(B-V)$) and age ($\tau$); and a second population of outliers, characterized by an offset from the isochrone ($Y_B$), a Gaussian variance of that offset ($V_B$), and a free parameter ($f$) that can capture other offsets such as systematics, inaccurate photometry, or areas of the CMD that the model grid insufficiently captures. From this analysis, we obtained an age estimate of $59^{+6}_{-5}$\,Myr and a reddening $E(B-V)=0.044^{+0.022}_{-0.024}$. 
To ensure the robustness of our search, we also evaluated the isochronal age at a 30\,pc and 50\,pc search radius, both arriving at the same age within $1\sigma$ of the 40\,pc search radius. With regards to the potential eclipsing binary discussed in \S~\ref{subsec:rot}, removing \textit{Gaia} DR3 2033585105381921024 from the isochronal fitting process did not noticeably change the resulting posteriors.

\begin{figure}[htb!]
    \centering
    \includegraphics[width=0.97\linewidth]{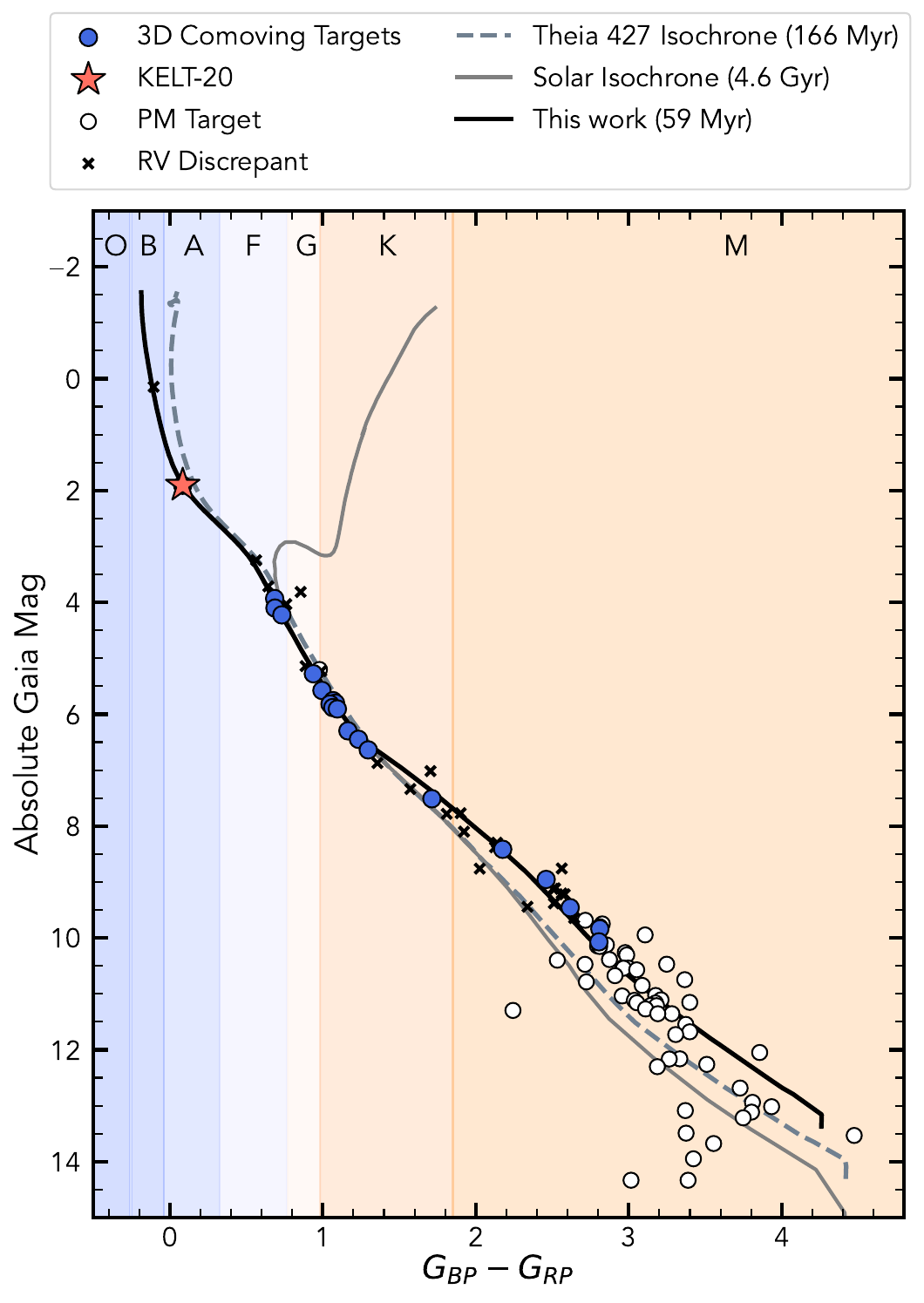}
    \caption{Color–magnitude diagram of comoving targets to \name{}. 3D comoving stars are shown as blue circles, while non-3D PM targets are shown as white circles. KELT-20 (coral star) is highlighted, along with a solid line showing the best-fit \texttt{PARSEC} isochrone derived from the isochronal analysis, corresponding to the association's median age. The dashed line shows a solar-age, solar-metallicity isochrone for comparison. Targets whose radial velocity differed by more than 3\,km\,s$^{-1}$ and thus not included in our analysis are shown as an `x.' Shaded background panels denote spectral type color regions based on \citet{Harre2021}.}
    \label{fig:CMD}
\end{figure}

In Figure~\ref{fig:CMD}, we show our best isochronal fit and the CMD positions of the PM and 3D comoving targets with KELT-20 displayed as a coral star. A population of still converging M-dwarfs is clearly observable in both the PM and 3D target sample. Also shown for context is a 4.6\,Gyr isochrone of solar metallicity and an isochrone corresponding to the reported age and extinction of Theia 427, which includes \name{} as a potential member (\citealt{KounkelCovey2020}; Theia 427 is discussed further in Section~\ref{sec:comparison}).

Lower-mass stars spend a significant amount of time on the pMS, making them useful benchmarks for estimating contamination in young clusters. 
To quantify the contamination rate of the PM and 3D comoving targets, we restricted the analysis to the color range $G_\mathrm{BP}-G_\mathrm{RP}=2.0$--$3.0$\,mag, which corresponds to the low-mass end of the 3D kinematic sample. 
Within this range, contaminants were defined as sources with CMD positions offset by more than $5\sigma$ from our isochronal age estimate, while non-contaminants were those within the $\pm5\sigma$ interval. 
Applying this criterion to the $2.0$--$3.0$\,mag subset, we find that 8 of 23 PM targets (${\sim}35\%$) are contaminants. 
Among the five 3D targets in the same color range, only one lies outside the $5\sigma$ bound, corresponding to a ${\sim}20\%$ contamination fraction, though the small sample size limits the statistical significance of this estimate.
We further assess potential contamination using gyrochronology in Section~\ref{subsec:rot}.

\subsection{Rotation Sequence Age Estimation}\label{subsec:rot}
Gyrochronology provides a robust and complementary method for estimating the age of coeval populations. 
This technique relies on the empirically established relationship between a star's age, rotation period, and mass. The age-dependent increase in stellar rotation period was first observed by \cite{Skumanich1972} in an analysis of stellar clusters with well-determined ages, where the rotation period was found to scale with the square root of the star's age. Subsequent studies have refined this relationship using rotational data from young clusters and field stars with ages determined by other methods, such as asteroseismology. 

The physical mechanism underpinning gyrochronology is widely understood to be magnetic braking, wherein angular momentum is lost as a result of the interaction between the stellar magnetic field and charged particles in the stellar wind \citep{Mestel1968, Gossage2023}. This interaction exerts a torque that gradually slows stellar rotation. The spin-down rate is influenced by both stellar mass and age, with lower-mass G- and K-type dwarfs generally experiencing faster rotational deceleration due to their deeper convective zones, which enhance magnetic coupling and accelerate angular momentum loss.

We analyzed the time series photometry of our PM and 3D comoving targets using data from the Transiting Exoplanet Survey Satellite (\textit{TESS}, \citealt{Ricker2015}) and the \textit{Kepler} mission \citep{Borucki2010}. 
For 64 targets, no \texttt{MAST} data were available \citep{MAST}.
We omitted \name{} from this analysis as it lies above the Kraft break, where rotation periods are challenging to measure and uninformative of age.
We measured rotation periods for 12 comoving targets, including one PM target and 11 3D comoving targets. The target \textit{Gaia} DR3 2033585105381921024 was removed, as it appears to be an eclipsing binary in the light curve. One other target (\textit{Gaia} DR3 2034082874897544448) was excluded due to contamination from nearby targets in the light curves, leaving ten rotation periods. Our rotational analysis methodology, along with the individual measured rotation periods, is discussed in Appendix~\ref{App:LC}.

\begin{figure*}[htb!]
    \centering
    \includegraphics[width=0.98\textwidth]{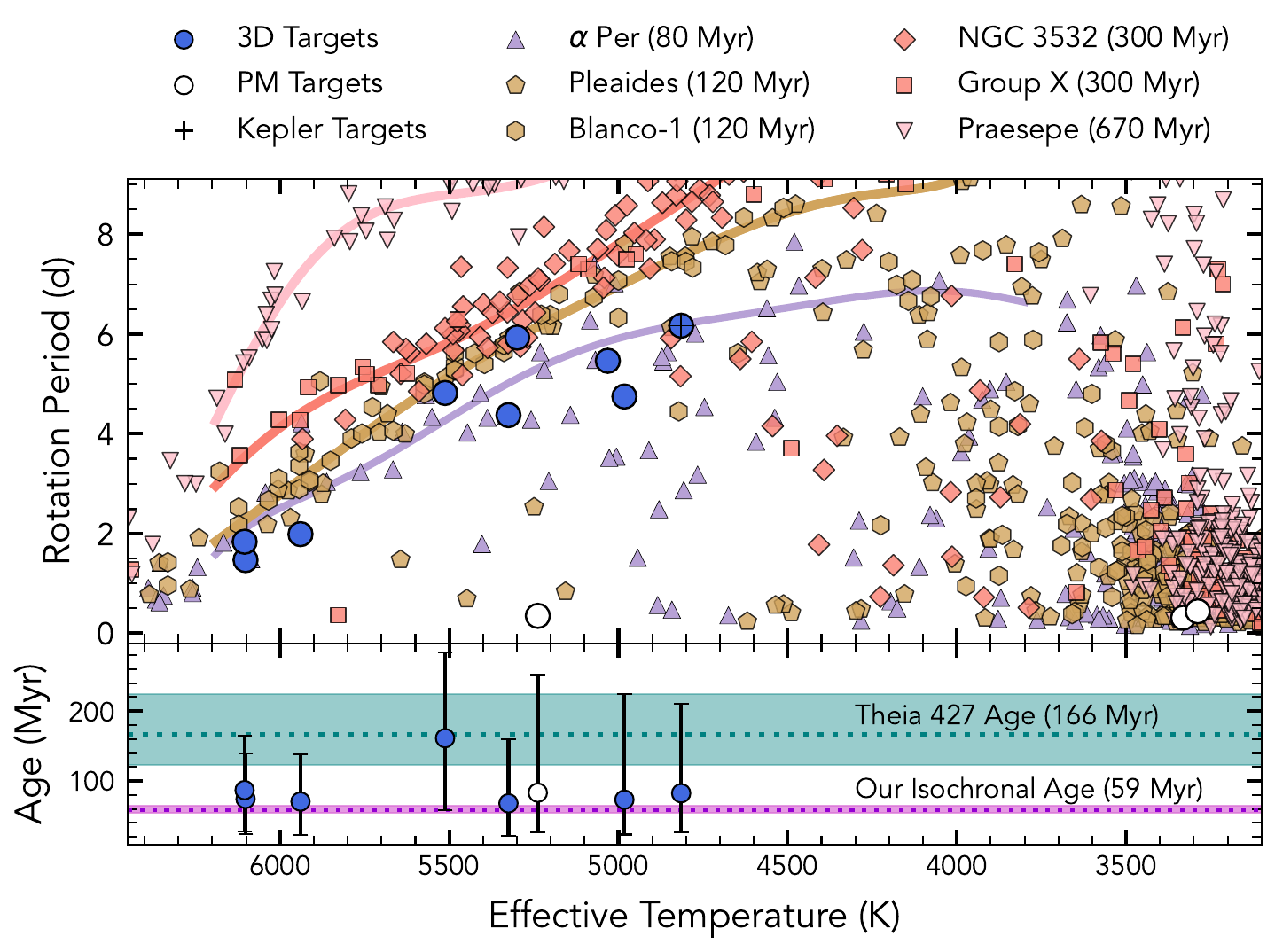}    
       \caption{Top Panel: rotation sequence for the comoving targets of \name{}. The target with \textit{Kepler} photometry is marked with a $+$. For comparison, we also show the 80\,Myr $\alpha$~Per association (purple triangles; \citealt{Boyle2023}), 120\,Myr Pleiades (yellow pentagons; \citealt{Rebull2016}), the 120\,Myr Blanco-1 (yellow hexagons; \citealt{Gillen2020}),the 300\,Myr Group X (salmon squares; \citealt{Messina2022}), the 300\,Myr NGC~3532 (salmon diamonds; \citealt{Fritzewski2021}), and the 670\,Myr Praesepe cluster (pink inverted triangles; \citealt{Rampalli2021}).  The corresponding gyrochrones for each association are also shown \citep{Bouma2023}. 
       Bottom Panel: Our \texttt{gyro-interp} results for the comoving targets displayed in the rotation sequence. The lower and upper errors are the 16th and 84th percentiles of the resulting probability distribution function. Also shown for comparison are the age of the Theia~427 moving group (teal band) and our isochronal age estimate (pink band), with shaded regions indicating the $1\sigma$ uncertainties.}
    \label{fig:rotationcombined}
    \end{figure*}

We constructed the effective temperature--rotation sequence for comoving targets to \name{}, as shown in the top panel of Figure~\ref{fig:rotationcombined}.
This sequence is contrasted with established gyrochronological benchmark systems, including $\alpha$~Per (80\,Myr, \citealt{Boyle2023}), the Pleiades (120\,Myr, \citealt{Rebull2016}), Blanco-1 (120\,Myr, \citealt{Gillen2020}), Group X (300\,Myr, \citealt{Messina2022}), NGC 3532 (300\,Myr, \citealt{Fritzewski2021}), and  Praesepe (670\,Myr, \citealt{Rampalli2021}).
The rotation sequence for the \name{} comoving targets closely follows that of $\alpha$~Per, providing additional support for the group's cohesiveness and youth.

To quantitatively estimate stellar ages from the derived rotation periods, we employed the tool \texttt{gyro-interp}, which uses a star's rotation period and effective temperature to predict the gyrochronological age by interpolating between the rotation sequences of known stellar associations \citep{Bouma2023}. As inputs, we used the \textit{Gaia} DR3 effective temperatures and our measured photometric rotation periods, along with a uniform 100\,K and $10\%$ error on the effective temperature and rotation period, respectively. \\ 

Gyrochronology provides another probe of the system age, though it is important to note that 80\,Myr is the youngest calibrated benchmark for \texttt{gyro-interp} ($\alpha$~Per; \citealt[][]{Bouma2023}). Applying \texttt{gyro-interp} to our sample, seven of the eight targets yielded ages younger than 90\,Myr, as shown in the bottom panel of Figure~\ref{fig:rotationcombined}. To combine these results, we performed the process of maximum likelihood for the ages provided by \texttt{gyro-interp}, assuming asymmetric Gaussians for the uncertainties. This yielded a population age of $84^{+32}_{-24}$\,Myr. However, because \texttt{gyro-interp} is not calibrated for ages below $\sim$80\,Myr, this value should be treated as an upper limit on the true system age rather than a precise measurement.

We next use gyrochronology as an independent means of estimating the contamination rate among \name{}'s comoving targets. The single proper-motion target cannot be assessed in this way, however, it appears young and is consistent with the isochronal age estimate. All 3D comoving targets fall within one sigma of the isochronal age, suggesting a low contamination rate for the group.

\subsection{Gaia Excess Variability}
Stars are much more magnetically active at young ages \citep{Lehtinen2016} and this activity often manifests as enhanced photometric variability, which \textit{Gaia} can capture as elevated flux uncertainties \citep{Rielleo2021, Thao2024}. 
The Excess Variability-based Age (EVA) framework \citep{Barber2023}\footnote{\url{https://github.com/madysonb/EVA}} leverages this variability–age correlation by using \textit{Gaia} DR3 ensemble variability statistics to infer group ages.
EVA constrains the population age rather than individual stellar ages, so no per-star EVA ages are reported.
We applied EVA to the full sample of \totalpm{} PM and 3D comoving targets of \name{}, obtaining a system age measurement of $48^{+17}_{-11}$\,Myr.
This value agrees well with the isochronal estimate but is younger than the gyrochronological age. However, this discrepancy is expected, as the \texttt{gyro-interp} tool does not provide reliable ages below $\sim$80\,Myr, and thus cannot fully capture the youth indicated by EVA. This strengthens the conclusion that this is a genuinely young ensemble.

\subsection{Investigating Youth with the Ca II Infrared Triplet}
Young stars also possess distinct spectroscopic features due to their increased magnetic activity. Quantifying the increased activity in the $H_\alpha$ \citep{Cincunegui2007}, Ca\,II H and K \citep{Wilson1968, Linsky1979, Mittag2013}, and Ca\,II infrared triplet \citep[IRT;][]{Martin2017} lines can provide valuable evidence for a star's age \citep{Lachaume1999, Newton2017, Kiman2021}. 

\begin{figure}[htb!]
     \centering
     \includegraphics[width=0.98\linewidth]{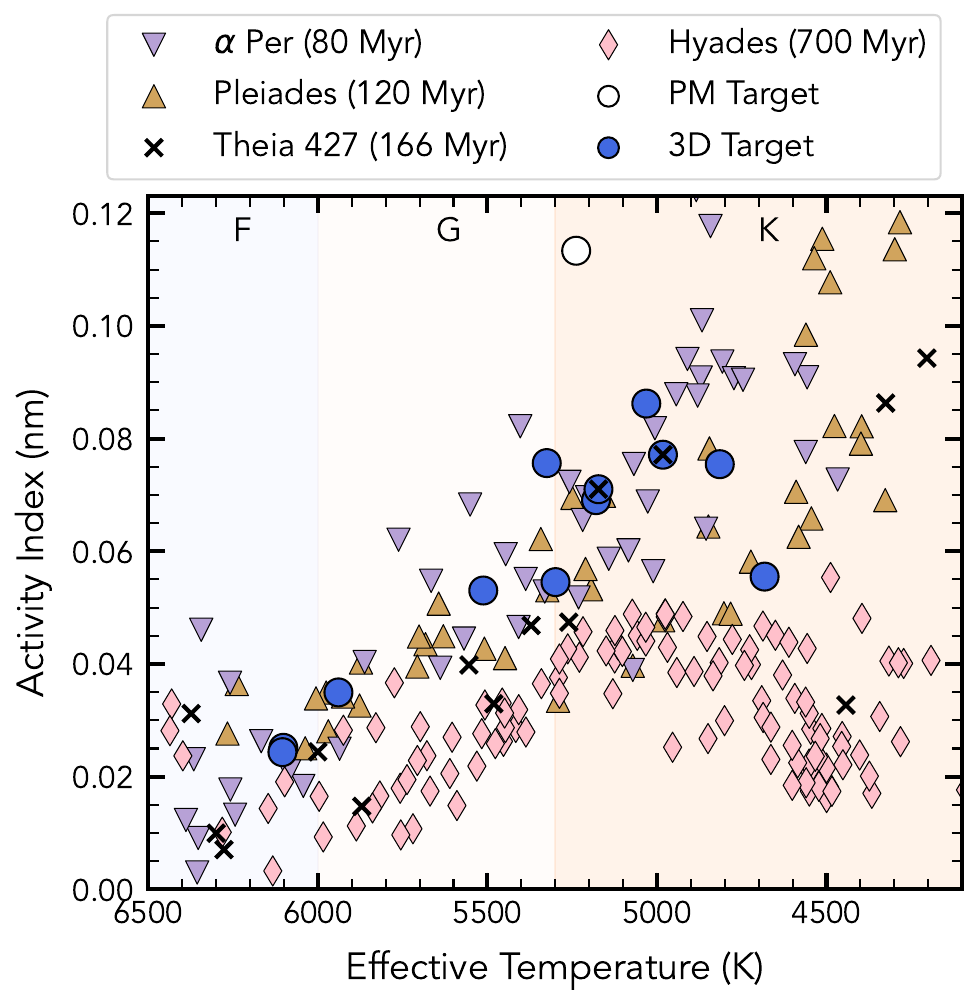}
     \caption{\textit{Gaia} DR3 activity index as a function of effective temperature. Shown are PM comoving targets (white circles), 3D comoving targets (blue circles), $\alpha$ Per (purple inverted triangles; \citealt{Boyle2023}), the Pleiades (yellow triangles; \citealt{Rebull2016}), and Hyades (pink diamonds; \citealt{Oh2020}). Theia~427 \citep{KounkelCovey2020} is also shown as black x's.}
     \label{fig:ACT}
 \end{figure}

To quantify the activity of the 3D comoving targets, we utilized the activity index provided by \textit{Gaia} DR3, which averages the excess equivalent width factor across each Ca II IRT line when compared to an inactive template spectrum. In Figure~\ref{fig:ACT}, we plot this activity as a function of \textit{Gaia} DR3 effective temperature for comoving targets along with other benchmark clusters. We find the overall activity of the comoving members is comparable to the younger $\alpha$ Per (80\,Myr) system as compared to the older Hyades (700\,Myr) system, although still relatively comparable to the slightly older Pleiades (120\,Myr). Measured activity indices are also shown for Theia 427 (166\,Myr), which appears to show more scatter and demonstrates the purity of our sample. The distribution of our comoving targets lying above the Hyades again highlights the low contamination rate of our sample.

\subsection{Combined Age Estimate}
Table~\ref{tab:Ages} lists the derived values, including isochrone fitting, gyrochronology, and EVA analysis from the ensemble. Although each method carries distinct assumptions and uncertainties, their overall consistency lends confidence to the results. For the combined analysis, we omit the gyrochronology estimate, as the youngest cluster on which \texttt{gyro-interp} has been validated is 80\,Myr. 
To combine the ensemble-derived age estimates, we treated each as a probability distribution with asymmetric uncertainties. These were modeled as asymmetric Gaussians, preserving the different upper and lower error bars. We then performed the process of maximum likelihood to find the combined age estimate from the photometric variability and isochronal constraints. From this analysis, we calculate an age of \age{}, which we adopt as the primary value for this work.
\begin{deluxetable}{lcccccccc}
\centering
\tabletypesize{\scriptsize}
\tablewidth{60pt}
\tablehead{\colhead{Method} & \colhead{Age (Myr)}
}
\startdata
Isochronal Fit (\texttt{PARSEC}) & $59^{+6}_{-5}$ \\
Variability (\texttt{EVA}) & $48^{+17}_{-11}$ \\
Gyrochronology (\texttt{gyro-interp}) & $84_{-24}^{+32}$ \\
\enddata
\caption{Summary of age estimates for \name{}. Isochrone fitting, gyrochronology, and variability-based ages are derived from the comoving ensemble.}
\label{tab:Ages}
\end{deluxetable}

\section{Comparison to Past Works}\label{sec:comparison}
Our analysis resulted in a much younger age for \name{} than prior works, such as \cite{Talens2018} and \cite{KounkelCovey2020}, which resulted in ages of $200^{+100}_{-50}$\,Myr and $166^{+58}_{-43}$\,Myr, respectively. 
To explain this age discrepancy, we first consider the methods used to derive the aforementioned results and then the possible sources of divergence.

\subsection{Theia 427 Moving Group}
\begin{figure}[!htbp]
    \centering
    \includegraphics[width=0.95\linewidth]{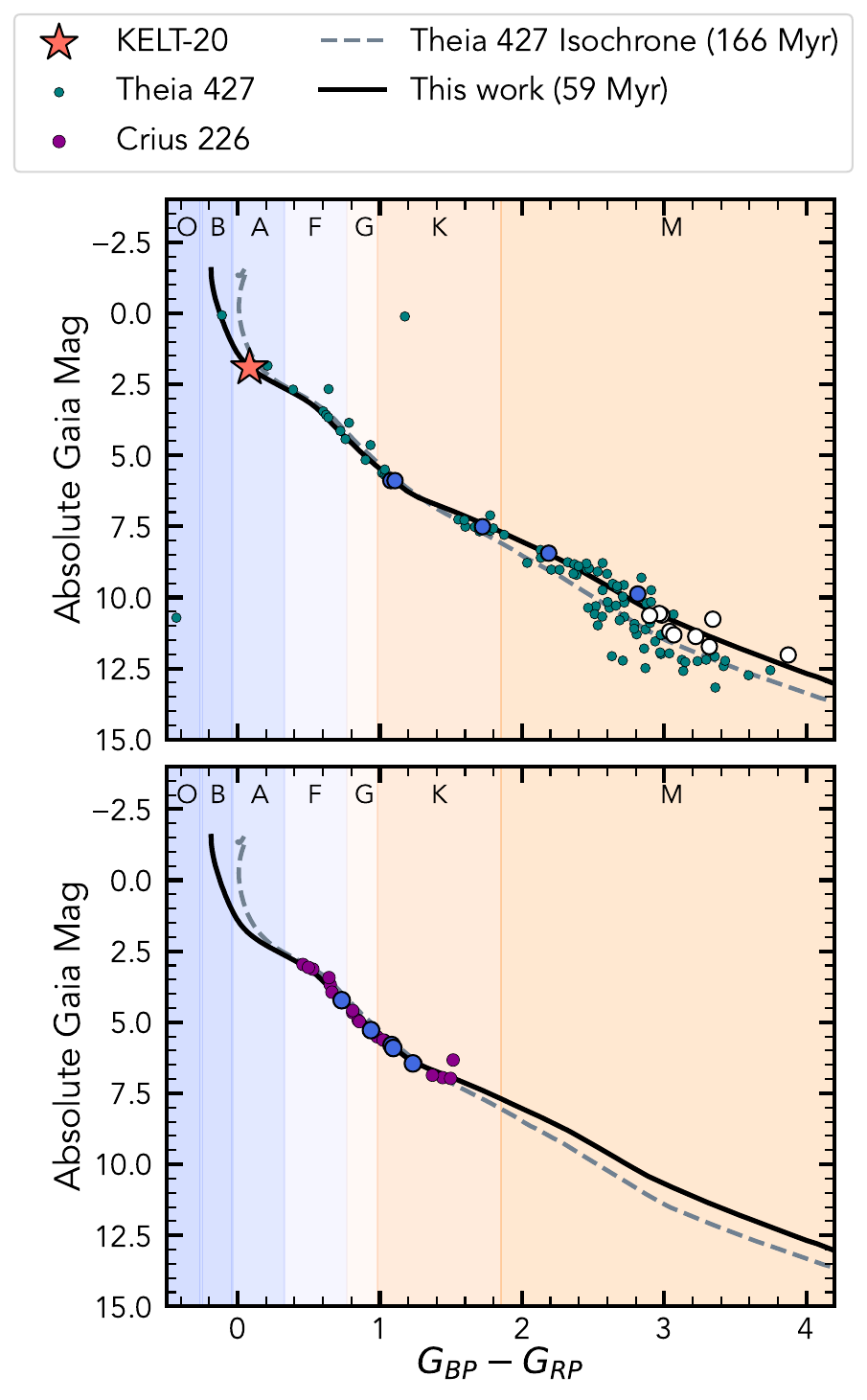}
     \caption{Top Panel: CMD of Theia~427 moving group (teal circles; \citealt{Kounkel2019, KounkelCovey2020}), showing overlapping PM targets (white circles) and 3D comoving targets (blue circles) present in both catalogs.
     Bottom Panel: CMD displaying the Crius~226 moving group (purple circles; \citealt{Moranta2022}), with common members and the same isochrones overlaid for comparison.}
     \label{fig:Comp}
 \end{figure}

\cite{KounkelCovey2020} identified comoving populations by applying reduced kinematics, incorporating galactic longitude, latitude, parallax, and proper motions.
Since this work predates the release of the \textit{Gaia} DR3 catalog, radial velocity measurements were not available for the majority of sources. This reduced the dimensionality of the analysis, increasing the potential for contamination. As shown in the top panel of Figure~\ref{fig:Comp}, there is an overlap of 15 sources between the comoving targets of \name{} analyzed in this work and Theia 427, of which \name{} was cited to be a member \citep{KounkelCovey2020}. 
Theia~427 contains a population of pMS M-dwarfs that are poorly fit by the isochronal age derived in their work.
Our interpretation is that the inclusion of older, converged MS M-dwarfs in the sample influenced the fit, resulting in an older age estimate than the one obtained in our analysis.
 
\subsection{Crius~226 Moving Group}
A subsample of our comoving targets also overlapped with the Crius~226 moving group \citep{Moranta2022}. In total, there were five common sources out of the 22 total Crius~226 population. According to \cite{Moranta2022}, Crius~226 is a subset of Group~59 \citep{Oh2017} and Theia~209 \citep{Kounkel2019, KounkelCovey2020}, with some overlap with Theia~133. They adopted the age of 160\,Myr reported for Theia~209 by \cite{KounkelCovey2020} as the age of Crius~226.
As shown in the bottom panel of Figure~\ref{fig:Comp}, the CMD position of the Crius~226 members is unlikely to enable useful constraints on isochronal age of the system. 
We applied EVA to the Crius~226 members reported by \cite{Moranta2022} and measured an age of $58_{-10}^{+13}$, consistent with that of our comoving targets.
Our result highlights the importance of re-evaluating ages with multiple age diagnostic techniques.
Our catalog includes only a subset of Crius~226 members, as we deliberately restrict our search to a 40\,pc radius centered on \name{} in order to reduce contamination from field stars. This conservative choice reflects our goal of deriving a high-fidelity age for \name{}, rather than constructing a complete census of a moving group.
We presume that Crius~226 is coeval with our set of comoving targets.

\section{Implications for KELT-20}\label{sec:Kelt-20b}
\name{} (HD\,185603, TOI-1151, TIC\,69679391, MASCARA-2) is an A2V star known to host a transiting hot Jupiter (HJ) companion \citep{Lund2017,Talens2018}. 
The system is located at a distance of \dist{}\,pc, resulting in a bright ($V = 7.6$\,mag) host star that is amenable to follow-up observations.
It is worth noting that the photometric and spectroscopic analyses performed by \citet{Talens2018} also resulted in a reasonably young (but less precise) age estimate for the star ($\tau=200^{+100}_{-50}$\,Myr), which overlaps with the moving group age estimate determined in this work.

The planetary companion, \pname{}, has a radius of 1.7\,\RJ{}  and orbits its host with a period of \porb{}\,d.
Despite extensive RV observations \citep{Lund2017, Talens2018}, the planet mass remains poorly constrained. More specifically, the spectral lines of the rapidly-rotating ($v\sin{(i)}=120\,\mathrm{km} \ \mathrm{s}^{-1}$) host star are significantly rotationally broadened, enabling only an upper limit of the companion mass (\mass{}\,\MJ{}, \citealt{Lund2017}). 

\paragraph{Atmospheric Evolution of KELT-20}\label{subsec:atmosevol}

As a HJ orbiting a bright star, \pname{} is an important test case for investigations of atmospheric physics under extreme conditions. 
A-type stars emit significant amounts of extreme-ultraviolet (XUV) radiation, which are enhanced at early stages of stellar evolution.
Therefore, planets like \pname{} experience extreme levels stellar irradiation, which can lead to radius inflation \citep{2012ApJ...751...59P} and atmospheric mass loss over time \citep{owen}. 
This extreme irradiation also results in strong temperature gradients and thermal dissociation of molecules in the atmosphere. 

It is therefore no surprise that the upper atmosphere of \pname{} has been examined extensively through high-resolution transmission spectroscopy \citep{Barris2018,Fu2022,Petz2023}. 
These investigations have led to detections of multiple atomic and molecular species that suggest the presence of a complex and dynamic exosphere influenced by intense stellar irradiation.
Prior studies have reported the presence of Na\,I and $\mathrm{H\alpha}$ \citep{Barris2018}, as well as detections of H$_2$O and CO \citep{Fu2022}. Additionally, the detection of Fe\,I and a tentative detection of Ni\,I have been suggested \citep{Petz2023}. Studies have also shown that the atmosphere of \pname{} extends significantly beyond the planet's photometric radius \citep{2021A&A...649A..29R}. 
These observations suggest that the upper atmosphere of the planet is inflated, likely due to the extreme stellar irradiation and strong atmospheric dynamics. 

The combination of extreme irradiation, potential atmospheric outflows, and indications of an extended atmosphere make \pname{} an ideal target to study how HJ atmospheres evolve, lose mass, and possibly transform in size over time. 
The age estimate of \pname{} provided in this work, which is both refined from prior estimates and robust in terms of the methodology, offers important new context for ongoing characterization efforts of this system.
Understanding these processes during the early stages of planetary evolution places places tight constraints on exoplanet atmospheric evolutionary models and helps contextualize the long-term fate of close-orbiting gas giants \citep{Dai2023, Karalis2025}.

\paragraph{Dynamical Evolution of \pname{}}\label{subsec:dynamical}

The dynamical history of HJs is a topic of active research. Two primary mechanisms for HJ migration have been proposed: high-eccentricity migration and disk-driven migration. 
High-eccentricity migration is believed to be driven by planet-planet scattering events \citep{Rasio1996}; secular chaos in multiplanet systems \citep{Teyssandier2019}; and Kozai-Lidov oscillations \citep{Kozai1962, Lidov1962, Naoz2011}, which can be induced by other planets \citep{Petrovich2015b} or stellar-mass objects \citep{Hamers2017, Wu2003}. 
This pathway can sometimes result in a significant stellar obliquity (misalignment between the stellar spin axis and planetary orbital plane, \citealt{Vick_2023}) and can operate over long timescales that are dependent on underlying tidal dissipation mechanisms. In contrast, disk-driven migration occurs during the planet’s early formation stages, requiring migration within the protoplanetary disk through interactions with the disk material \citep{Goldreich1980, Lin1986}.

Observational trends suggest that high-eccentricity migration may dominate the current HJ population, as alignment trends differ for stars above and below the Kraft break. Stars with significant convective envelopes (below the Kraft break) tend to exhibit well-aligned systems ($\lambda \approx 0$), while more massive stars with radiative envelopes (above the Kraft break) display a much wider range of obliquities \citep{Winn2010}.

Given its age constraints and orbital configuration, \pname{} presents an interesting test case. With the occurrence rate of giant planets appearing to peak around A-stars \citep{Wolthoff2022} but with a smaller HJ occurrence rate compared to less massive stars \citep{Beleznay2022}, \pname{} represents a compelling test case in understanding migration mechanisms around more massive stars. This system is shown to be well-aligned \citep{Lund2017}, with a sky-projected obliquity of $\lambda = 3.9\pm 1.1^\circ$ \citep{Singh2024}.
Further, we report that \pname{} is in a circular orbit given the presence of a secondary transit eclipse at half the orbital period. 

\begin{figure}
    \centering
    \includegraphics[width=0.95\linewidth]{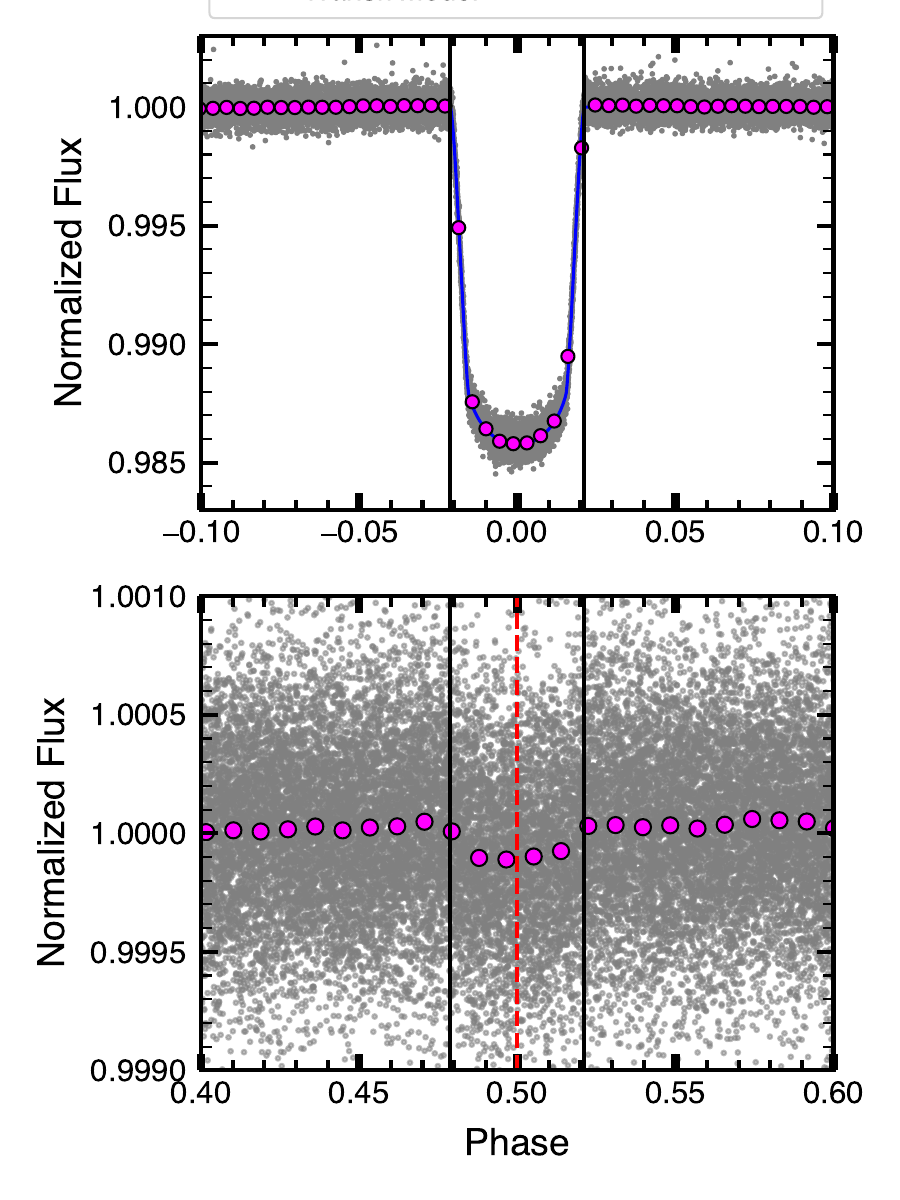}
    \caption{Phase-folded \textit{TESS} curves from the transit fitting process. Top panel: phase-folded light curve of the transit of KELT-20 (grey points), along with the binned data (purple points) and the transit model (blue curve). The black vertical lines display the $t_1$ and $t_4$ contact points of the transit. Bottom panel: phase-folded TESS light curve for the secondary eclipse of \pname{}. The red dashed line dictates the location of $P/2$, and the black vertical lines dictate the $t_1$ and $t_4$ contact points of the transit.}
    \label{fig:transit}
\end{figure}

More specifically, we observed this signature in an analysis performed on the TESS light curves using the differential MCMC code \texttt{edmcmc}\footnote{\url{https://github.com/avanderburg/edmcmc}} and the transit-fitting package \texttt{batman}.\footnote{\url{https://lkreidberg.github.io/batman}}
We fit for the semi-major axis $(a)$, orbital period $(P)$, time of inferior conjunction $(t_0)$, planetary radius $(R_p)$, and quadratic limb-darkening parameters ($u_1$, $u_2$). We ran our transit fitting software for a total of 5000 links to ensure the walkers were well-mixed. We adopted the Gelman-Rubin statistic for our convergence criteria, requiring a value $<1.05$ for every parameter. After fitting for the orbital period, we see a clear secondary eclipse at $t_0 + P/2$, implying a likely circular orbit. Another possible, but unlikely scenario is that we are viewing an eccentric orbit that happens to have the longitude of periastron at $\approx 0^\circ$, however, in this configuration the transit and eclipse would have different durations. The transit fit and \textit{TESS} data are shown in Figure~\ref{fig:transit}.

The well-aligned and circular nature of \pname{}'s orbit has some potentially important dynamical implications, especially when paired with the age constraints of the system. Below, we briefly discuss how this informs the plausibility of major migration mechanisms.
\begin{itemize}
    \item \textbf{Disk-driven Migration:}
    
   Disk-driven migration operates while the disk is still present, which typically takes place within the first $\sim$10\,Myr \citep{Heller2019}. This is further enhanced by the longer disk lifetimes promoted by the low extreme-UV and X-ray flux from its host A-type star \citep{Nakatani2021}. Further, simulations predict that planetary migration of planetary embryos around intermediate-mass stars may be most efficient at $\approx 1.7\,M_\odot$, although formation of HJs in simulations around higher mass stars remains challenging \citep{Johnston2024}. Thus, this pathway could feasibly explain the planet's orbital configuration within the timescale constrained by the system's age. 

    \item \textbf{Obliquity-inducing Migration Mechanisms:} 
    Given the low obliquity of the system, migration mechanisms dependent on obliquity excitation must be dampened on timescales similar to the age of the system. The stellar obliquity dampening timescale, as determined by the equilibrium tide model, is
    \begin{equation}
        \tau_\textrm{RE} = (0.25 \times 5 \times 10^9\,\textrm{yr}) q^{-2} (1+q)^{-5/6} \bigg(\frac{a/R_\star}{6}\bigg)^{\frac{17}{2}},
    \end{equation}
    where $\tau_{\textrm{RE}}$ is the timescale for obliquity damping with a radiative envelope, $q= M_p / M_\star$ is the planet to star mass ratio, $a$ is the semi-major axis of the planet, and $R_\star$ is the radius of the host star \citep{Albrecht2012}.
    
    Using \cite{Lund2017} for the orbital parameters, we calculate an obliquity dampening timescale of $10^{17}$\,yr, which is ten orders of magnitude longer than the current age of the system. 
    While resonance locking has also been proposed as an obliquity-dampening mechanism, given the reliance on g-mode coupling, it has been shown to be inefficient for hot stars like \name{} \citep{Zanazzi2024}.
    
    However, there are some important caveats to bear in mind. 
    Firstly, \name{} lacks an accurate inclination measurement.
    Thus, \pname{} might be misaligned with respect to the \textit{true} obliquity ($\psi$), although this is unlikely given how close the projected obliquity happens to be to $0^{\circ}$.
    In addition, the obliquity might have been excited during the pre-main sequence phase. During that time, the star would have been almost entirely convective \citep{Palla1993, Baraffe2010}, enabling more efficient obliquity damping. 
    While outside the scope of this work, a more in-depth analysis of the dynamical evolution of the system, particularly at early timescales, is a critical next step of investigation. 
 
    \item \textbf{Eccentricity-inducing Migration Mechanisms:}
    This formation mechanism would need to account for the circular orbital configuration of \pname{}. 
    The timescale for circularization is,
    \begin{equation}
    \tau_\mathrm{circ} = \frac{2}{81} \frac{Q_p'}{n} \frac{M_p}{M_*} \bigg(\frac{R_p}{a}\bigg)^{-5} F_p^{-1}\,,
    \end{equation}
    where $\tau_\textrm{circ}$ is the circularization timescale, $G$ is the gravitational constant, and $Q_p'$ is the modified tidal quality factor of the planet ($Q_p'=Q_p/k_2$; \citealt{Matsumura2008}). Here, $k_2$ represents the Love number of the planet and $n$ is the mean motion $(n=\sqrt{G(M_*+M_p)/a^3})$. Here, we assume that the planet and the star are tidally locked. $F_p$ is defined as 
    \begin{equation}
        F_p(e) = f_1(e) - \frac{11}{18} f_2(e)\,,
    \end{equation}
    with 
    \begin{equation}
    f_1(e) = (1 + \frac{15}{4} e^2 + \frac{15}{8} e^4 + \frac{5}{64}e^6 )/(1-e^2)^{13/2}\,,
    \end{equation}
    \begin{equation}
        f_2(e) = (1 + \frac{3}{2} e^2 + \frac{1}{8} e^4)/(1-e^2)^{5}\,.
    \end{equation}
    Rather than calculating the circularization timescale, one can instead set $\tau_\mathrm{circ} = 60$\,Myr constraint and calculate the maximum $Q_p'$ needed to result in the current-day circular orbit. Using the parameters from \cite{Lund2017}, we calculated $Q_p'$ for $M_p = 1,2,3$\,\MJ{}. In these cases, we calculate a range of $Q_p' \le 10^{7}$, which is reasonable for a Jupiter-like planet \citep{Goldreich1966, Rice2022}. However, gas giant formation theory indicates that the HJ should have originally formed at a significantly larger semi-major axis. 
    As shown in Figure~\ref{fig:Qp_plot}, if \pname{} started  $\sim 1$\,AU with $M_p = 1$\,\MJ, this would imply a $Q_p' < 10$, which is unphysical for a nominal gas giant. 
\begin{figure}[htb!]
        \centering
        \includegraphics[width=0.98\linewidth]{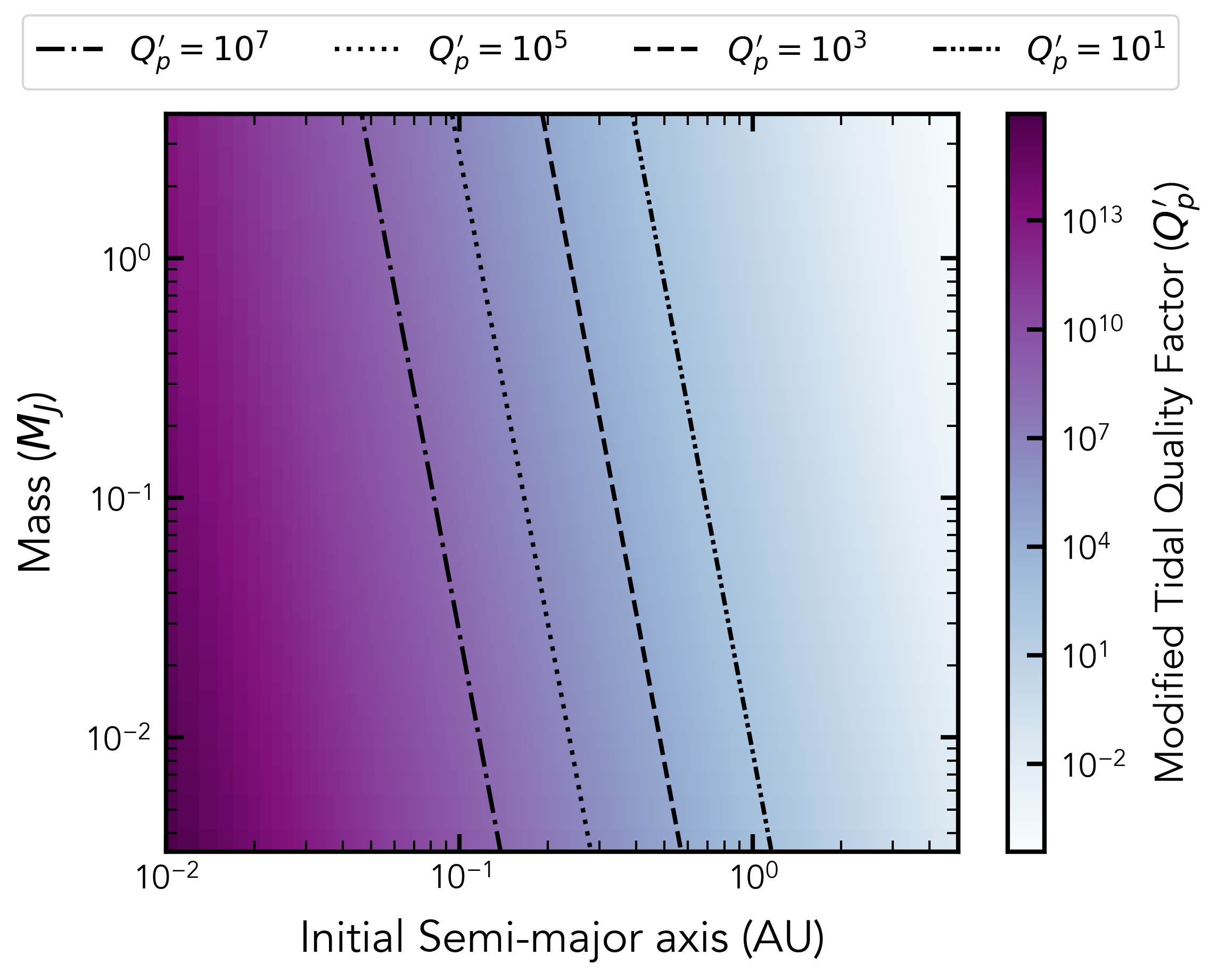}
        \caption{The tidal modified quality factor ($Q_p'$) as a function of initial semi-major axis and mass for the \pname{} system. Lines of constant $Q_p'$ are also shown, with values $10$, $10^3$, $10^5$, and $10^7$. }
        \label{fig:Qp_plot}
    \end{figure}
    
    It is important to note that estimates of $Q_p$ values may differ for extremely young planets. 
    At such a close orbit and young age, \pname{} is expected to be quite inflated independent of the incident stellar radiation from its host star and its internal entropy from planet formation \citep{Burrows2000, Tremblin2017, Sarkis2021}. 
    This may imply a complex and radically different interior structure to the closest solar system analog of Jupiter, leading to fundamentally different tidal dissipation processes. Thus, while it is unlikely, it is challenging to fully rule out high-eccentricity migration.   
  
\end{itemize}    

The dynamical history of \pname{} need not occur through one primary pathway, but could instead be a combination of mechanisms leading to its observed configuration. The interplay between disk-driven migration and high-eccentricity migration in young moving groups need not be a simple process, especially in consideration of the system's rapidly evolving properties due to its young age. Further, assumptions about the tidal dissipation factor $Q_p$ and also considering the tides raised on a contracting host star might have significantly influenced \pname{}'s orbital evolution.

\section{Summary}\label{sec:summary}
In this work, we derive a high-fidelity age for \name{} of \age{} by combining multiple, independent ensemble diagnostics from a 40\,pc, volume-limited, comoving sample. 
Our kinematic search identified \totalpm{} proper–motion candidates, of which \totaltd{} possessed \textit{Gaia} DR3 radial velocities enabling full 3D confirmation. 
Multiple independent diagnostics, including gyrochronological age estimates, indicate minimal contamination within our young, comoving sample.  

This refined age is noticably younger than prior estimates: $166^{+58}_{-43}$\,Myr from Theia~427 \citep{KounkelCovey2020} and  $200^{+100}_{-50}$\,Myr from stellar models of \name{} by \citet{Talens2018}. This difference arises from our stricter 3D comovement selection and conservative spatial cut, which was designed to limit contamination, rather than to recover a complete census of potential moving group members.  

\pname{} stands out as a well-aligned HJ orbiting a young A-type star. Its circular orbit and lack of significant stellar obliquity \citep{Singh2024} raise major questions about its migration history. Our measurement expands the small sample of young star–planet systems with precise age estimates. As one of the few known young ultra–hot Jupiters transiting a bright star, this system is ideally suited for transmission spectroscopy, offering a rare chance to probe atmospheric composition and mass loss at early epochs. In addition, its orbital architecture provides a useful test case for studying the dynamical pathways that shape HJs during their formative stages.

\begin{acknowledgments}
AD gratefully acknowledges the generous support of the \textit{Peter Livingston Scholars Program}, whose contributions to undergraduate research have played a key role in the development of this work.
This material is based upon work supported
by the National Science Foundation Graduate Research Fellowship under Grant No. DGE-2140743. Any opinion, findings, and conclusions or recommendations expressed in this material are
those of the authors(s) and do not necessarily reflect the views of the National Science
Foundation.
Support for this research was provided by the Office of the Vice Chancellor for Research and Graduate Education at the University of Wisconsin--Madison with funding from the Wisconsin Alumni Research Foundation.
This paper includes data collected by the \tess{} mission, which are publicly available from the Mikulski Archive for Space Telescopes (MAST). Funding for the \tess{} mission is provided by NASA’s Science Mission Directorate.
This research has made use of the Exoplanet Follow-up Observation Program (ExoFOP; DOI: 10.26134/ExoFOP5) website, which is operated by the California Institute of Technology, under contract with the National Aeronautics and Space Administration under the Exoplanet Exploration Program. 
This work has made use of data from the European Space Agency (ESA) mission \emph{Gaia},\footnote{\url{https://www.cosmos.esa.int/gaia}} processed by the \emph{Gaia} Data Processing and Analysis Consortium (DPAC).\footnote{\url{https://www.cosmos.esa.int/web/gaia/dpac/consortium}} 
This research has made use of the VizieR catalogue access tool, CDS, Strasbourg, France. The original description of the VizieR service was published in A\&AS 143, 23. 
\end{acknowledgments}

\facilities{
GALEX \citep{Martin2005}; 
Gaia EDR3 and DR3 \citep{GaiaMission, gaiadr3}; 
Kepler \citep{Borucki2010}; 
Mikulski Archive for Space Telescopes \citep{MAST};
Pan-STARRS\,1 \citep{Chambers2016};
\tess{} \citep{Ricker2015};
Two Micron All-Sky Survey \citep{Skrutskie2006}}

\software{
\texttt{astroquery} \citep{astroquery},
\texttt{batman} \citep{batman}, 
\texttt{EAGLES} \citep{eagles,Weaver2024},
\texttt{edmcmc} \citep{vanderburgedmcmc}, 
\texttt{EXOFASTv2} \citep{Eastman2013}, 
\texttt{Lightkurve} \citep{lightkurve},
\texttt{matplotlib} \citep{matplotlib},
\texttt{PAdova and TRieste Stellar Evolution Code} \citep{parsec2012}, 
\texttt{PyAstronomy} \citep{pya},
\texttt{Scipy} \citep{scipy},
\texttt{SIESTA} \citep{siesta},
\texttt{wdwarfdate} \citep{Kiman2022}
}

\bibliographystyle{aasjournalv7}
\bibliography{biblio.bib}

\begin{appendix}

\section{Light Curve Analysis}\label{App:LC}
To determine the rotation periods of the comoving targets, we made use of archival photometric time series data. 
The Transiting Exoplanet Survey Satellite (\tess{}; \citealt{Ricker2015}) provides broad sky coverage, dividing the celestial sphere into observation sectors, each covering a $24^\circ$ by $96^\circ$ field of view. The spacecraft focuses on one sector at a time, observing continuously for about 27 days before moving to the next sector. 

Whenever available, we used light curves from the TESS-SPOC pipeline \citep{2020RNAAS...4..201C}, prioritizing short cadence data. 
When these TESS-SPOC data were unavailable, we used 30-minute cadence SAP light curves from the MIT Quick Look Pipeline (QLP; \citealt{QLP, QLP2}), and data from the TESS Asteroseismic Science Operations Center (TASOC) pipeline \citep{2021AJ....162..170H, 2021ApJS..257...53L}.
For one target, we used \textit{Kepler/K2} mission data \citep{Borucki2010}, including light curves processed by community reduction tools \citep{Vanderburg2014, 2023AJ....166..265M}.
All the aforementioned datasets are publicly available through the Mikulski Archive for Space Telescopes (MAST; \citealt{MAST}).

Due to differing systematics across distinct \tess{} sectors and \textit{Kepler} quarters, we evaluated each individually. Using the software platform \texttt{Lightkurve} \citep{lightkurve}, we performed sigma clipping of outliers at the $5\sigma$ level, as these are often the result of systematics and stellar flares. To determine the level of light curve contamination, we relied on the TESS Input Catalog contamination ratio metric, which measures the amount of contamination originating from nearby targets \citep{Stassun2018}. As higher values indicate more contamination, we adopt an upper limit of 1 for the contamination ratio \citep{Boyle_2025}. We find that one target (Gaia DR3 2034082874897544448) does not meet this criteria. This is due to a brighter, nearby source that overlaps with the \textit{TESS} aperture. We mark all targets that passed all criteria as a "0" in the Quality Flag column of Tables~\ref{tab:subset_td} and ~\ref{tab:subset_pm}, and those that failed at least one of the criteria as a ``1."

\begin{figure*}[htb!]
\centering
    \includegraphics[width=0.82\linewidth]{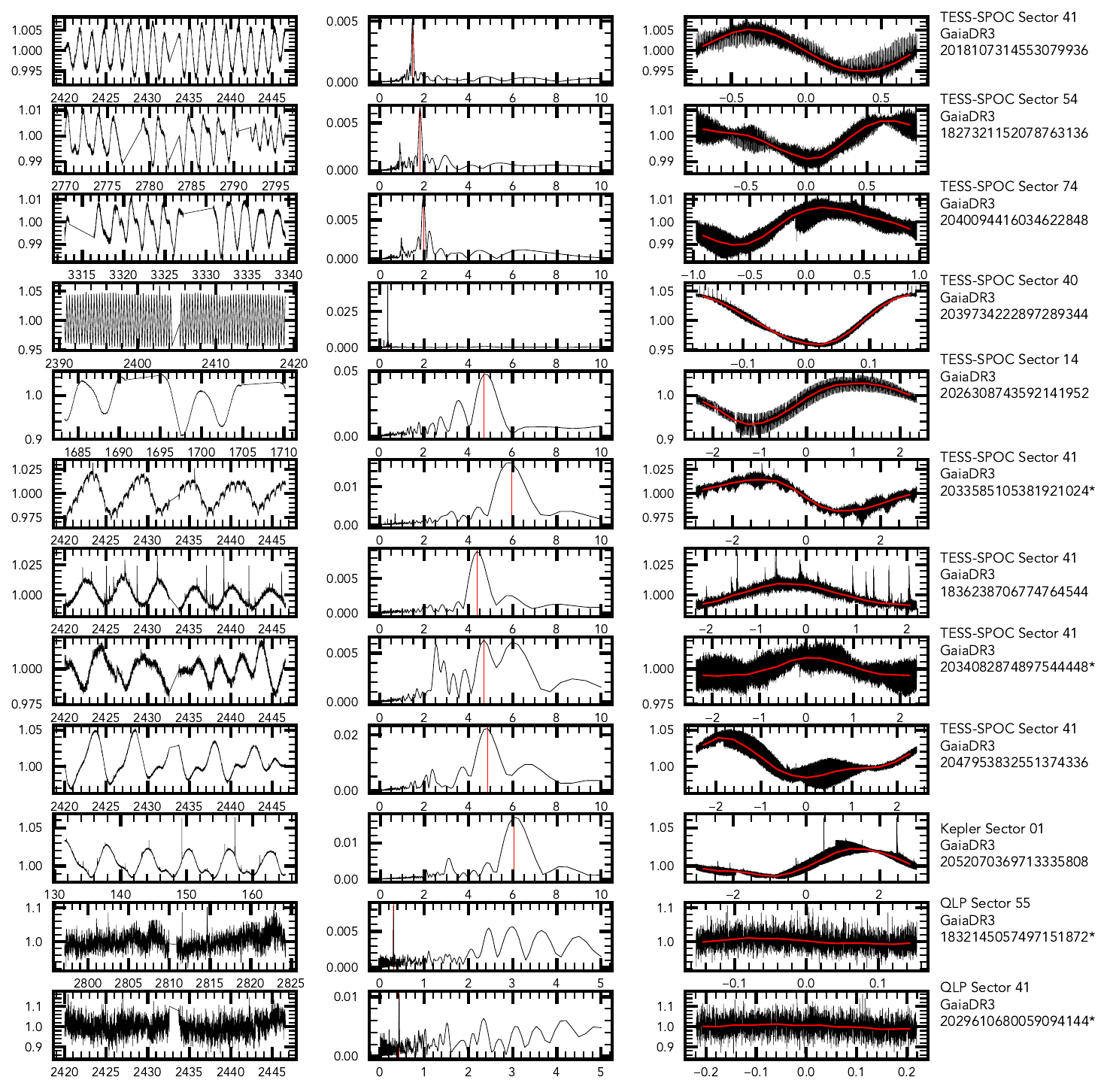}
    \caption{Results of the rotation analysis performed on comoving targets to \name{}. Left panel: light curve of the star as measured by \textit{TESS} and \textit{Kepler}. Middle Panel: Lomb-Scargle periodogram of the light curve. Right Panel: phase-folded light curve using the maximum power of the periodogram. An asterisk (*) denotes that the target was excluded from the \texttt{gyro-interp} analysis, either due to binarity, photometric aperture contamination, or effective temperatures outside the algorithm bounds.}
    \label{fig:TotalRots}
\end{figure*}

We ran a Lomb-Scargle \citep{Lomb1976, Scargle1982} periodogram analysis, searching for periodic signals with periods between $0.3-13$\,days. 
To determine the photometric rotation period of a star, we visually inspected the most prominent five peaks in the power spectra for each sector, along with their associated phase-folded light curves.
For targets with compelling periodic signals, we took the median of the values observed across all sectors as our reported rotation period. To check for pipeline systematics, we required the rotational signal to be present in at least two different pipelines. Example light curves, periodograms, and phase-folded light curves are displayed in Figure~\ref{fig:TotalRots}. For the one target (Gaia DR3 2052070369713335808, KIC 3128488) also observed by \textit{Kepler}, we observe the same rotation period between both \textit{Kepler} and \textit{TESS} light curves, adding verification to our data quality and analysis.

\section{Comoving Target Parameters}
In Tables~\ref{tab:subset_td} and \ref{tab:subset_pm}, we present comprehensive stellar parameters for our comoving target sample. Table~\ref{tab:subset_td} contains photometric and activity measurements for \name{} and the 19 confirmed 3D comoving companions identified through full space motion analysis. Table~\ref{tab:subset_pm} provides corresponding parameters for the 58 proper motion comoving candidates that share similar tangential proper motion velocities but lack sufficient radial velocity constraints for full 3D kinematic confirmation. These measurements include Gaia photometry, stellar effective temperatures, photometric rotation periods where detectable periodic variability is present, and gyrochronological age estimates derived from the rotation-age relation when both rotation periods and reliable stellar parameters are available. We also provide chromospheric activity indices and cross-matches with existing moving group catalogs to provide a complete characterization of the comoving population around \name{}. 

\begin{deluxetable*}{cccccccccccc}
\tabletypesize{\scriptsize}
\tablecaption{Parameters for \name{} and the 19 3D comoving targets in this study.\label{tab:subset_td}}
\tablehead{
\colhead{Index} & \colhead{Gaia DR3 ID} & \colhead{$M_G$} & \colhead{BP--RP} & \colhead{$T_{\mathrm{eff}}$} & \colhead{$P_{\mathrm{rot}}$} & \colhead{Quality Flag} & \colhead{Cont. Ratio} &\colhead{Gyro Age} & \colhead{Activity Index} & \colhead{Theia} & \colhead{Crius} \\
\colhead{} & \colhead{} & \colhead{(mag)} & \colhead{(mag)} & \colhead{(K)} & \colhead{(d)} &\colhead{} & \colhead{} &  \colhead{(Myr)} & \colhead{(nm)} & \colhead{} & \colhead{}
}
\startdata
1 & 2033123654092592384 & 1.91 & 0.08 & 8849 & -- & -- & -- & -- & -- & 427 & -- \\
2 & 2018107314553079936 & 3.93 & 0.69 & 6102 & 1.5 & 0 & 0.06 & 74 & 0.025 & -- & -- \\
3 & 1827321152078763136 & 4.10 & 0.69 & 6105 & 1.8 & 0 & 0.07 & 87 & 0.024 & 165 & -- \\
4 & 2040094416034622848 & 4.22 & 0.73 & 5939 & 2.0 & 0 & 0.04 & 70 & 0.035 & 209 & 226 \\
5 & 2026308743592141952 & 5.28 & 0.94 & 5512 & 4.8 & 0 & 0.55 & 161 & 0.053 & 165 & 226 \\
6 & 2033585105381921024 & 5.57 & 0.99 & 5299 & 5.9 & 1 & 0.63 & -- & 0.055 & -- & -- \\
7 & 1836238706774764544 & 5.76 & 1.07 & 5325 & 4.4 & 0 & 0.15 & 68 & 0.076 & -- & -- \\
8 & 2034082874897544448 & 5.80 & 1.08 & 5031 & 5.5 & 1 & 1.98 & -- & 0.086 & -- & 226 \\
9 & 2072786882461493120 & 5.82 & 1.05 & 5180 & -- & -- & -- & -- & 0.069 & 133 & -- \\
10 & 2032986455663675264 & 5.88 & 1.06 & 5172 & -- & -- & -- & -- & 0.071 & 427 & -- \\
11 & 2047953832551374336 & 5.91 & 1.09 & 4982 & 4.7 & 0 & 0.26 & 73 & 0.077 & 427 & 226 \\
12 & 1858988946686228224 & 6.30 & 1.16 & -- & -- & -- & -- & -- & -- & -- & -- \\
13 & 2052070369713335808 & 6.45 & 1.23 & 4815 & 6.2 & 0 & 0.55 & 82 & 0.075 & 209 & 226 \\
14 & 4534695740349652480 & 6.64 & 1.30 & 4682 & -- & -- & -- & -- & 0.056 & -- & -- \\
15 & 2021873348046521856 & 7.51 & 1.71 & -- & -- & -- & -- & -- & -- & 427 & -- \\
16 & 1836495550127142912 & 8.42 & 2.18 & -- & -- & -- & -- & -- & -- & 427 & -- \\
17 & 2024060998210085632 & 8.95 & 2.46 & 3523 & -- & -- & -- & -- & -- & 209 & -- \\
18 & 2043431816087540864 & 9.46 & 2.62 & 3446 & -- & -- & -- & -- & -- & 209 & -- \\
19 & 2028502990831600000 & 9.84 & 2.81 & 3476 & -- & -- & -- & -- & -- & 427 & -- \\
20 & 2026101485642100608 & 10.07 & 2.81 & 3364 & -- & -- & -- & -- & -- & 209 & -- \\
\enddata
\tablecomments{
The complete catalog of comoving targets, including additional parameters, is available via VizieR. The first row corresponds to \name{}.\\
Column descriptions: (1) {Index} --- row number from 1 to 20. (2) {Gaia DR3 ID} --- unique identifier from the Gaia Data Release 3 catalog. (3) {$M_G$} --- absolute Gaia $G$-band magnitude (mag). (4) {BP--RP} --- Gaia color index (mag). (5) {$T_{\mathrm{eff}}$} --- stellar effective temperature (K) from Gaia DR3. (6) {$P_{\mathrm{rot}}$} --- photometric rotation period (days). (7) Quality Flag, 0 if passed all criteria, 1 if failed at least one criteria. (8) TIC Contamination Ratio. (9) {Gyro Age} --- stellar age estimate (Myr) from \texttt{gyro-interp}. (10) {Activity Index} --- Gaia DR3 activity index (nm). (11) {Theia} --- moving group identifier from the Theia catalog \citep{KounkelCovey2020}. (12) {Crius} --- moving group identifier from the Crius catalog \citep{Moranta2022}.
}
\end{deluxetable*}

\begin{deluxetable*}{cccccccccccc}
\tabletypesize{\scriptsize}
\tablecaption{Parameters for the 58 proper motion targets in this study.\label{tab:subset_pm}}
\tablehead{
\colhead{Index} & \colhead{Gaia DR3 ID} & \colhead{$M_G$} & \colhead{BP--RP} & \colhead{$T_{\mathrm{eff}}$} & \colhead{$P_{\mathrm{rot}}$} & \colhead{Quality Flag} & \colhead{Cont. Ratio} &\colhead{Gyro Age} & \colhead{Activity Index} & \colhead{Theia} & \colhead{Crius} \\
\colhead{} & \colhead{} & \colhead{(mag)} & \colhead{(mag)} & \colhead{(K)} & \colhead{(d)} &\colhead{} & \colhead{} &  \colhead{(Myr)} & \colhead{(nm)} & \colhead{} & \colhead{}
}
\startdata
1 & 2039734222897289344 & 5.20 & 0.98 & 5238 & 0.3 & 0 & 0.16 & 83 & 0.113 & 209 & -- \\
2 & 2033501783027661312 & 9.69 & 2.72 & 3377 & -- & -- & -- & -- & -- & -- & -- \\
3 & 2055678760727397248 & 9.75 & 2.83 & 3358 & -- & -- & -- & -- & -- & -- & -- \\
4 & 1832145057497151872 & 9.80 & 2.81 & 3331 & 0.3 & 0 & 0.34 & -- & -- & 300 & -- \\
5 & 2029610680059094144 & 9.95 & 3.11 & 3288 & 0.4 & 0 & 0.54 & -- & -- & -- & -- \\
6 & 2102296267021398784 & 10.12 & 2.85 & 3321 & -- & -- & -- & -- & -- & -- & -- \\
7 & 4520713152212097920 & 10.14 & 2.80 & 3392 & -- & -- & -- & -- & -- & -- & -- \\
8 & 2045367025294525824 & 10.15 & 2.81 & -- & -- & -- & -- & -- & -- & -- & -- \\
9 & 2047311271057377792 & 10.26 & 2.98 & -- & -- & -- & -- & -- & -- & 133 & -- \\
10 & 1816892185512187520 & 10.31 & 2.99 & 3277 & -- & -- & -- & -- & -- & 300 & -- \\
11 & 4532526854893200512 & 10.39 & 2.87 & 3324 & -- & -- & -- & -- & -- & -- & -- \\
12 & 2093533193707324928 & 10.40 & 2.53 & 3498 & -- & -- & -- & -- & -- & -- & -- \\
13 & 2043457792050251648 & 10.47 & 3.25 & 3312 & -- & -- & -- & -- & -- & -- & -- \\
14 & 2042795439373138176 & 10.47 & 2.71 & 3420 & -- & -- & -- & -- & -- & -- & -- \\
15 & 4588078030611966464 & 10.54 & 2.96 & 3339 & -- & -- & -- & -- & -- & -- & -- \\
16 & 2025187963300590976 & 10.54 & 3.00 & 3268 & -- & -- & -- & -- & -- & 427 & -- \\
17 & 2022748181343169408 & 10.54 & 2.97 & 3335 & -- & -- & -- & -- & -- & 427 & -- \\
18 & 1827104277746527360 & 10.57 & 3.05 & 3315 & -- & -- & -- & -- & -- & -- & -- \\
19 & 2024525954212276096 & 10.68 & 2.91 & 3345 & -- & -- & -- & -- & -- & 427 & -- \\
20 & 1836090414459033856 & 10.75 & 3.37 & 3277 & -- & -- & -- & -- & -- & 427 & -- \\
21 & 2031190575195261184 & 10.79 & 2.72 & 3383 & -- & -- & -- & -- & -- & 165 & -- \\
22 & 2047276567750640000 & 10.85 & 3.09 & 3278 & -- & -- & -- & -- & -- & -- & -- \\
23 & 2026054515872644864 & 11.03 & 3.18 & 3238 & -- & -- & -- & -- & -- & -- & -- \\
24 & 4520243866916459520 & 11.04 & 2.96 & -- & -- & -- & -- & -- & -- & -- & -- \\
25 & 2032790639509718144 & 11.11 & 3.21 & 3245 & -- & -- & -- & -- & -- & 209 & -- \\
26 & 1832488242563682176 & 11.12 & 3.04 & 3279 & -- & -- & -- & -- & -- & 209 & -- \\
27 & 2061906394598039168 & 11.15 & 3.40 & 3203 & -- & -- & -- & -- & -- & 133 & -- \\
28 & 2032553759138084864 & 11.16 & 3.05 & 3276 & -- & -- & -- & -- & -- & 427 & -- \\
29 & 2023184893615437568 & 11.16 & 3.17 & 3240 & -- & -- & -- & -- & -- & 209 & -- \\
30 & 2033585517698923648 & 11.21 & 3.15 & 3241 & -- & -- & -- & -- & -- & -- & -- \\
31 & 1864307245767302272 & 11.21 & 3.18 & 3218 & -- & -- & -- & -- & -- & -- & -- \\
32 & 2025184007577225088 & 11.27 & 3.11 & 3262 & -- & -- & -- & -- & -- & 427 & -- \\
33 & 2038517445804253440 & 11.30 & 2.24 & -- & -- & -- & -- & -- & -- & 209 & -- \\
34 & 2035235055680858880 & 11.36 & 3.28 & 3224 & -- & -- & -- & -- & -- & 427 & -- \\
35 & 2032953539032072832 & 11.36 & 3.19 & 3224 & -- & -- & -- & -- & -- & 209 & -- \\
36 & 2020095987464484224 & 11.55 & 3.37 & -- & -- & -- & -- & -- & -- & 165 & -- \\
37 & 2030049690136077056 & 11.68 & 3.40 & 3115 & -- & -- & -- & -- & -- & 427 & -- \\
38 & 2078432462354988416 & 11.73 & 3.31 & 3174 & -- & -- & -- & -- & -- & 133 & -- \\
39 & 2044328841475774464 & 12.05 & 3.86 & -- & -- & -- & -- & -- & -- & 427 & -- \\
40 & 2036237161762062976 & 12.16 & 3.33 & 3119 & -- & -- & -- & -- & -- & 209 & -- \\
41 & 2048899167725394176 & 12.17 & 3.27 & -- & -- & -- & -- & -- & -- & -- & -- \\
42 & 2042340928752139648 & 12.26 & 3.51 & 3128 & -- & -- & -- & -- & -- & -- & -- \\
43 & 4515336639123732864 & 12.30 & 3.19 & 3175 & -- & -- & -- & -- & -- & -- & -- \\
44 & 2020095918737679872 & 12.57 & -- & -- & -- & -- & -- & -- & -- & -- & -- \\
45 & 2047311855196557440 & 12.68 & -- & -- & -- & -- & -- & -- & -- & -- & -- \\
46 & 1818260111114158720 & 12.69 & 3.73 & 3024 & -- & -- & -- & -- & -- & -- & -- \\
47 & 2035372430212899072 & 12.94 & 3.81 & 2840 & -- & -- & -- & -- & -- & -- & -- \\
48 & 2060168474705864064 & 13.02 & 3.93 & -- & -- & -- & -- & -- & -- & -- & -- \\
49 & 4318766504390517760 & 13.09 & 3.37 & 3009 & -- & -- & -- & -- & -- & -- & -- \\
50 & 2030050549118777984 & 13.12 & 3.80 & -- & -- & -- & -- & -- & -- & -- & -- \\
51 & 1834518692632659840 & 13.22 & 3.75 & -- & -- & -- & -- & -- & -- & -- & -- \\
52 & 1870502684554459520 & 13.49 & 3.38 & -- & -- & -- & -- & -- & -- & -- & -- \\
53 & 2029830754178590336 & 13.51 & -- & -- & -- & -- & -- & -- & -- & -- & -- \\
54 & 2035893736170572416 & 13.53 & 4.47 & 2629 & -- & -- & -- & -- & -- & -- & -- \\
55 & 2020865332062208512 & 13.68 & 3.56 & -- & -- & -- & -- & -- & -- & -- & -- \\
56 & 2059760109222646912 & 13.95 & 3.42 & -- & -- & -- & -- & -- & -- & -- & -- \\
57 & 2102546715154285952 & 14.33 & 3.02 & -- & -- & -- & -- & -- & -- & -- & -- \\
58 & 2077320134540441856 & 14.33 & 3.39 & -- & -- & -- & -- & -- & -- & -- & -- \\
\enddata
\tablecomments{
The complete catalog of comoving targets, including additional parameters, is available via VizieR.\\
Column descriptions: (1) {Index} --- row number from 1 to 58. (2) {Gaia DR3 ID} --- unique identifier from the Gaia Data Release 3 catalog. (3) {$M_G$} --- absolute Gaia $G$-band magnitude (mag). (4) {BP--RP} --- Gaia color index (mag). (5) {$T_{\mathrm{eff}}$} --- stellar effective temperature (K) from Gaia DR3. (6) {$P_{\mathrm{rot}}$} --- photometric rotation period (days). (7) Quality Flag, 0 if passed all criteria, 1 if failed at least one criteria. (8) TIC Contamination Ratio. (9) {Gyro Age} --- stellar age estimate (Myr) from \texttt{gyro-interp}. (10) {Activity Index} --- Gaia DR3 activity index (nm). (11) {Theia} --- moving group identifier from the Theia catalog \citep{KounkelCovey2020}. (12) {Crius} --- moving group identifier from the Crius catalog \citep{Moranta2022}.
}
\end{deluxetable*}

\end{appendix}
\end{document}